\documentclass[aps,pre,reprint,bibnotes,footinbib,showkeys,groupedaddress,superscriptaddress]{revtex4-1}
\usepackage{graphicx}
\usepackage{siunitx}
\usepackage{amsmath,amsfonts,amssymb,amsthm}
\usepackage{verbatim}

\usepackage{natbib}
\usepackage{color}
\definecolor{darkblue}{rgb}{0,0,0.5}
\definecolor{darkred}{rgb}{0.5,0,0}
\usepackage[colorlinks=true,urlcolor=darkblue,citecolor=darkblue,linkcolor=darkred,hyperfootnotes=false]{hyperref}
\usepackage{mathbbol}
\usepackage[normalem]{ ulem }
\usepackage{soul}

\DeclareSymbolFontAlphabet{\amsmathbb}{AMSb}

\newtheorem{theorem}{Theorem}

\theoremstyle{plain} 
\newcommand{\thistheoremname}{}
\newtheorem{genericthm}[theorem]{\thistheoremname}

\newcommand{\M}[1]{\mathcal{#1}}
\newcommand{\bo}[1]{\mathbf{#1}}
\newcommand{\bos}[1]{\boldsymbol{#1}}

\newcommand{\h}[1]{\hat{#1}}
\newcommand{\tl}[1]{\tilde{#1}}

\newcommand{\mb}[1]{\mathbb{#1}}
\newcommand{\ti}{\text{i}}

\newcommand{\tdir}{\tilde{\mathcal{N}}_R}

\newcommand{\ecoli}{\textit{E. coli}~}

 \definecolor{GREEN}{rgb}{0.3,.54,0}

\begin{document}

\title{Single-Trajectory Characterization of Active Swimmers in a Flow.}

\author{Gaspard Junot}
\affiliation{Laboratoire PMMH-ESPCI Paris, PSL Research University, Sorbonne Universit\'e and Denis Diderot, 7, quai Saint-Bernard, Paris, France}

\author{Eric Cl\'ement}
\affiliation{Laboratoire PMMH-ESPCI Paris, PSL Research University, Sorbonne Universit\'e and Denis Diderot, 7, quai Saint-Bernard, Paris, France}
\affiliation{Institut Universitaire de France (IUF)}

\author{Harold Auradou}
\affiliation{Universit\'e Paris-Saclay, CNRS, FAST, 91405, Orsay, France}

\author{Reinaldo  Garc\'{i}a-Garc\'{i}a}
\email{reinaldomeister@gmail.com}
\affiliation{Laboratoire PMMH-ESPCI Paris, PSL Research University, Sorbonne Universit\'e and Denis Diderot, 7, quai Saint-Bernard, Paris, France}

\begin{abstract}	
We develop a maximum likelihood method to infer relevant physical properties of elongated active particles. Using individual trajectories of advected swimmers as input, we are able to accurately determine their rotational diffusion coefficients and an effective measure of their aspect ratio, also providing reliable estimators for the uncertainties of such quantities. We validate our theoretical construction using numerically generated active trajectories upon no-flow, simple shear, and Poiseuille flow, with excellent results.  Being designed to rely on single-particle data, our method eases applications in experimental conditions where swimmers exhibit a strong morphological diversity. We briefly discuss some of such ongoing experimental applications, specifically, in the characterization of swimming \ecoli in a flow.
 \end{abstract}

\maketitle
\section{Introduction}

To monitor passive probes, living cells and microorganisms moving in their environment, modern instrumentation is now offering possibilities for long-time tracking and trajectory reconstruction, with excellent spatial and temporal resolution \cite{Manzo2015}. The collected data sets contain in principle relevant informations on the internal and external processes, of either  mechanical or biological origin.  However, inferring them is still a challenging task which has  motivated the recent development of adapted analytical tools ~\cite{Sarfati2017,Seyrich2018,Garcia2018,Frishman2020,Grebenkov1,Grebenkov2}. 

Tracking colloids close to thermal equilibrium  in various media was proven quite fruitful as it allowed the development of a quantitative micro-rheology instrumentation \cite{Gardel2005}. This technique was further extended to investigate complex fluids~\cite{Abou2004,Puertas2014}, and dynamical processes inside living cells \cite{Wirtz2009,Struntz2009,Manzo2015}. In recent years, much effort was  dedicated to the study of passive colloidal particles as they diffuse in active suspensions composed of self-propelling entities~\cite{Wu2000,Mino2011,Jeanneret2016}. Several studies focus on the contribution of the active environment to the diffusion of tracers~\cite{Chen2007, Mino2013,Kasyap2014}, while others aim at generalizing the notion of thermal bath and to characterize the thermodynamic fluctuations associated with the passive particle in complex environments~\cite{Marconi2015,Fodor2016,Dabelow2019, Burkholder2019}.
However, there has been much less focus on understanding how an active particle gets itself affected by the properties of its the environment.

In this work, we develop a maximum likelihood (ML) approach to study elongated active swimmers using raw data of swimming tracks. The general aim is to extract from a stochastic data set, the physical parameters that characterize the particle dynamics as well as the source of rotational noise. In general, such data encompass the active transport contribution resulting from the interaction between the flow and the swimmer, and the stochastic component associated with rotational noise. 
Although our framework is general, we focus here on three particular situations, $(i)$ a free swimmer, $(ii)$ a swimmer in a shear flow and, finally, $(iii)$ a swimmer in a Poiseuille flow.
We take into account inherent limitations encountered in experiments, such as the finite duration of the tracks and their discrete sampling. The method is first tested against numerical simulations, hence showing an excellent agreement between prescribed parameter values and our estimators.

Our tool has been conceived to study \ecoli mutants (smooth swimmers) in a parabolic flow, however, it can be generalized to different experimental situations by adapting the underlying stochastic process to the system at hand. 
For instance, many studies focused on the behavior of synthetic microswimmers in dynamic environments~\cite{katuri2016artificial}, such as externally imposed chemical gradients~\cite{baraban2013chemotactic}, or flows~\cite{palacci2015artificial}. Our method can also be used to extract key parameters from trajectories of synthetic microswimmers in such situations.

Smooth swimmers were recently considered in Ref.~\cite{Junot2019}. It was shown that a deterministic model, consisting of an advected swimming ellipsoid, fairly reproduces bacterial trajectories at short times. However, at longer times, a stochastic component comes into play  as a multiplicative noise leading to the stochastic exploration of the phase space. Using the ML method, we are able to extract the effective rotational diffusivity from the swimming trajectories, which can be compared quantitatively with the Brownian diffusion of an ellipsoid in a viscous fluid.

ML approaches have been applied earlier to the characterization of either passive or active tracers. For instance, in Ref.~\cite{C8CP04043E} a combination of Bayesian and ML analysis was used to infer the stochastic model that fits best a given single-passive-particle track. In Ref.~\cite{Masson1802} similar methods led to a noninvasive protocol to infer molecular chemotactic responses from bacterial trajectories. Regarding those studies, the main differences with our method are: $(i)$ it disentangles the deterministic (advective) and random (diffusive) components of single bacterial trajectories to better understand how active swimmers interact with different flow profiles, $(ii)$ the input of our ML procedure are raw dynamical trajectories (i.e., the set of positions and orientation vectors of the particle as function of time).

The paper is organized as follows. In Sec.~\ref{sec:prem} we introduce the theoretical model and briefly discuss the general philosophy of our ML method. In Sec.~\ref{sec:ML} we sketch the steps to build the log-likelihood and derive our ML estimators as well as their uncertainties to leading order in the number of sampling points. In Sec.~\ref{sec:validation} we use numerical simulations to validate our method in three configurations, no-flow, simple shear flow, and Poiseuille flow. A preliminary application of our method in the analysis of experimental tracking data is discussed in Sec.~\ref{sec:experimental}, and general conclusions are given in Sec.~\ref{sec:discussion}. All technical details regarding our calculations and experiments are left to the Appendix.

\section{Preliminaries}
\label{sec:prem}

\subsection{Stochastic dynamics of the active Betherton-Jeffery model}

In Ref.~\cite{Junot2019}, it was shown that rotational diffusion considerably affects the trajectories of mutant \ecoli in a flow.
This fact represents a strong motivation to consider the effects of rotational noise in the dynamics of the orientation vector of smooth swimmers.
We focus on the model studied in Ref.~\cite{Junot2019} in presence of noise, which we refer to as the stochastic active Betherton-Jefferey (SABJ) model from now on. The model describes the behavior of an ellipsoid swimming at a constant speed in a flow. Its dynamics read:
\begin{align}
\label{eq-motion-r}
\dot{\bo r} &=v_0\bo{p}+\bo{v}^F(\bo r),\\
\label{eq-motion-p}
\dot{\bo p} &= \big(\mb 1-\bo p\otimes\bo p\big)\big[\beta\mb E(\bo r)+\mb \Omega(\bo r)\big]\bo p-2D_R\bo p\nonumber\\
 &+\sqrt{2D_R}\bo p\wedge\bos \xi_R.
\end{align}

Above, $v_0$ is the self-propulsion velocity of the particle, $\bo v^F$ is the local flow velocity, the symbol $\mb 1$ denotes the identity matrix, and $\otimes$ is used to denote tensor products. The number $\beta=(r^2-1)/(r^2+1)$ is the Betherton parameter, which represents a measure of the geometrical asymmetry of the swimmer ($r$ is the aspect-ratio of the ellipsoidal particle). 
The components of the tensors $\mb E$ and $\mb \Omega$ are given as follows, $E_{ij}(\bo r)=(\partial_{x_i}v^F_j+\partial_{x_j}v^F_i)/2$ and $\Omega_{ij}(\bo r)=(\partial_{x_i}v^F_j-\partial_{x_j}v^F_i)/2$.  

In Eq.~\eqref{eq-motion-p}, rotational diffusion is encoded in the diffusion coefficient $D_R$ and the Gaussian white noise $\bos \xi_R$ which has zero mean and variance $\langle\bos \xi_R(t)\otimes\bos \xi_R(t')\rangle=\mb 1\delta(t-t')$. 
In this work, Eq.~\eqref{eq-motion-p} is interpreted in the Ito sense. The Ito term $-2D_R\bo p$ is thus needed to guarantee the conservation of the norm of $\bo p$.
Note that, although one could also introduce a translational diffusion term in Eq.~\eqref{eq-motion-r}, such contribution can be assumed negligible in typical experimental situations. 
Consider, for instance, an \ecoli bacterium which is few microns length, swimming at a typical speed of $\SI{25}{\micro\meter\per\second}$. The distance over which it diffuses in $\SI{1}{\second}$ is $\approx \SI{0.1}{\micro\meter}$ to be compared to  the $\SI{25}{\micro\meter}$ travelled due to its activity.

\begin{figure}[t!]
	\centering
	\includegraphics[scale= 0.19]{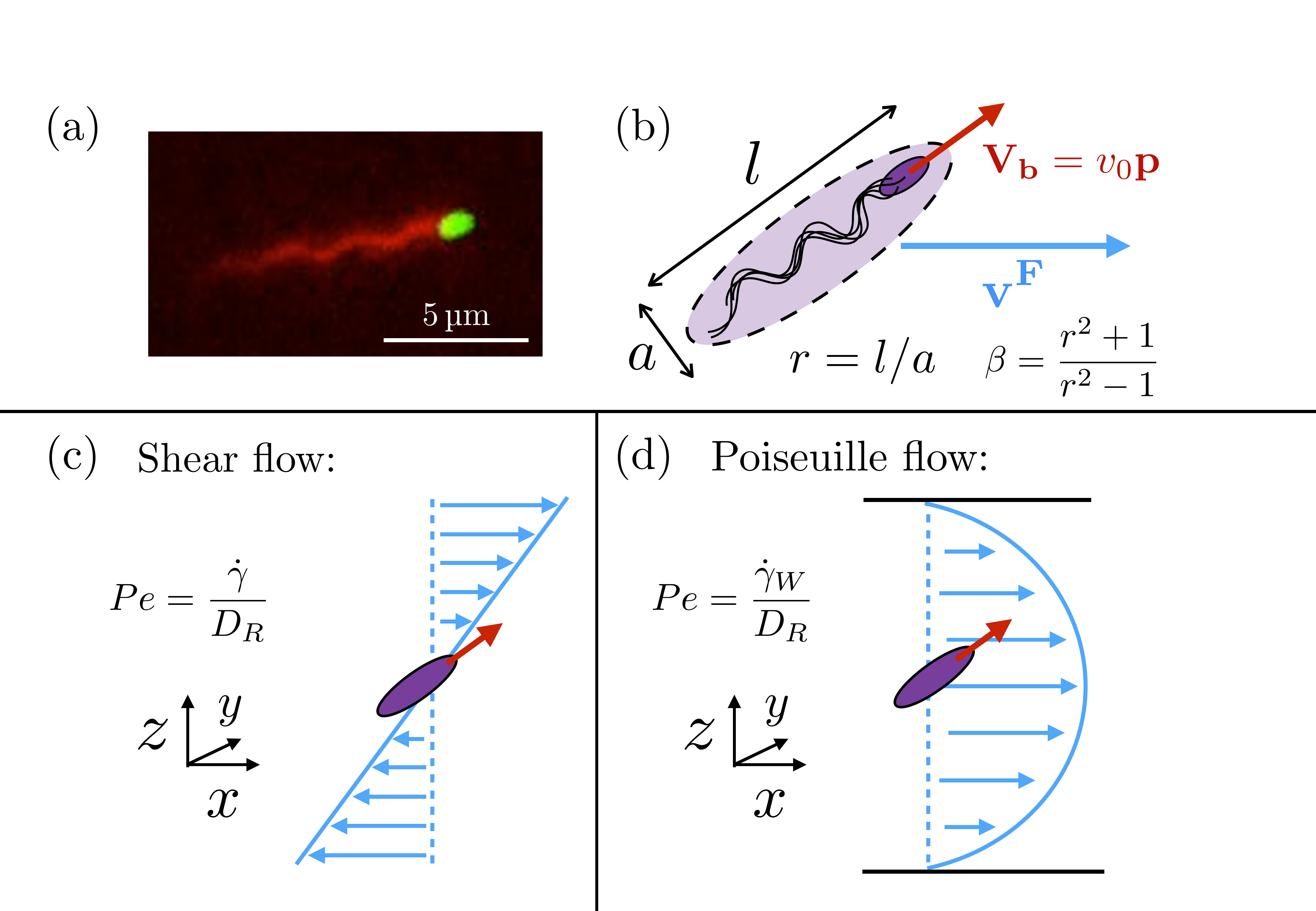}
	\vspace{0 mm}
	\caption{(a) Picture of an \ecoli bacterium with the head GFP labelled (green color) and the flagella bundle tagged in red with a specific alexa fluor marker. (b) Sketch of the effective ellipsoid of aspect ratio $r$ modeling the bacterium. (c) Simple shear flow geometry characterized by a constant shear rate $\dot{\gamma}$ and, (d) Poiseuille flow geometry; $\dot{\gamma}_W$ is the maximal shear rate. We also introduce the P\'eclet number, which is the natural dimensionless form of the inverse of the rotational diffusion coefficient.}
	\label{Fig_coord_BJ_v3}
\end{figure}

In Fig.~\ref{Fig_coord_BJ_v3} we illustrate our model system, making explicit the way in which geometrical parameters are defined. It is important to remark that the bacterium is a complex object and that the optical determination of $\beta$ is non-trivial. In other words, $\beta$ is an effective parameter and its determination based on imagery alone may not be reliable. However, for an elongated bacterium such as \ecoli its expected  value should be close to one (see the sketch in Fig.~\ref{Fig_coord_BJ_v3}(b)). In the same figure, we also show how the P\'eclet number is defined in presence of shear and Poiseulle flows. The P\'eclet number plays an important role in numerics because it is the natural dimensionless form of the inverse of the rotational diffusion coefficient, see Sec.~\ref{sec:validation} for more details.

An important role is played by the Markov kernel of the process~\eqref{eq-motion-r}-\eqref{eq-motion-p}, which we denote by $K(\bo r,\bo p,t|\bo r', \bo p',t')$. We focus here on stationary flow profiles in which case the propagtor only depends on $\Delta t=t-t'$ due to time-translation invariance. We thus simplify the notations by writing $K(\bo r,\bo p,t|\bo r', \bo p',t')=K(\bo r,\bo p,\Delta t|\bo r', \bo p',0)\equiv K(\bo r,\bo p,\Delta t|\bo r', \bo p')$. The propagator satisfies the Fokker-Planck equation $\partial_t K=\h L_{FP}K$ with initial condition $K(\bo r,\bo p,0|\bo r',\bo p')=\delta(\bo r-\bo r')\delta(\bo p-\bo p')$. The Fokker-Planck operator associated to these dynamics reads
\begin{align}
\label{FP-operator}
\h L_{FP}(\bo r,\bo p) = &D_R\sum_{i,j}\partial_{p_i,p_j}^2\cdot\tl M_{ij}(\bo p)-\sum_i\partial_{p_i}\cdot h_i(\bo r,\bo p)\nonumber\\
 &-\sum_i\partial_{x_i}\cdot v_i(\bo r,\bo p).
\end{align}

To write~\eqref{FP-operator}, we condensed the translational part of the motion into $\bo v(\bo r,\bo p)=v_0\bo p+\bo v^F(\bo r)$, and the drift term in Eq.~\eqref{eq-motion-p} into $\bo h(\bo r,\bo p)=(\mb 1-\bo p\otimes\bo p)(\beta\mb E(\bo r)+\mb \Omega(\bo r))\bo p-2D_R\bo p$. Additionally, each element of the matrix $\tl{\mb M}(\bo p)$ is a quadratic form in $\bo p$, $\tl M_{ij}(\bo p)=\sum_{l,q}T_{ijlq}p_{l}p_{q}$, where
\begin{equation}
\label{M-matrix}
T_{ijlq}=\sum_{k}\epsilon_{ilk}\epsilon_{jqk}=\delta_{ij}\delta_{lq}-\delta_{iq}\delta_{jl},
\end{equation}
and $\epsilon_{ijk}$ denotes the Levi-Civita symbol.

\subsection{Estimating parameters from single trajectories}

By studying individual trajectories of bacteria, one could access the statistics of $D_R$ and $\beta$ over a population. To achieve an accurate measurement of those parameters, we develop an ML method. We start by constructing the log-likelihood characterizing a single-bacterium trajectory; then the relevant physical parameters are determined by maximizing it. 

In this paragraph we sketch out the general philosophy of our procedure.
In experiments, the trajectory of a bacterium is recorded at regularly spaced time instants $\{t_\alpha\}$, such that $\Delta t_\alpha=t_{\alpha+1}-t_\alpha\equiv f^{-1}$ for each $\alpha=0,1,\ldots,N-1$, where $f$ is the sampling frequency. The data has the form of a discrete set of $N$ position and orientation values, $\{\Gamma\}$, where $\Gamma=(\bo r,\bo p)$. Assuming that the dynamics of the swimmer are compatible with Eqs.~\eqref{eq-motion-r}-\eqref{eq-motion-p}, the probability of measuring that particular track can be expressed as follows:
\begin{equation}
\label{track-proba}
\M P\big[\big\{\Gamma\big\}\big]=\prod_{\alpha=0}^{N-1}K(\Gamma(t_{\alpha+1}),f^{-1}|\Gamma(t_\alpha)).
\end{equation}

Although not made explicit, the probability of the track is parametrized by the relevant physical quantities of the model, i.e., $v_0$, $D_R$, and $\beta$. One can introduce the log-likelihood as $\M S=\ln\M P$, or more explicitly
\begin{equation}
\label{log-likelihood}
\M S=\sum_{\alpha=0}^{N-1}\ln K(\Gamma(t_{\alpha+1}),f^{-1}|\Gamma(t_\alpha)),
\end{equation}
anticipating the importance of the Markov kernel of the SABJ model in this work. By maximizing the log-likelihood with respect to the parameters, we obtain the best estimates compatible with the hypothesis that the data are generated by Eqs.~\eqref{eq-motion-r}-\eqref{eq-motion-p}.

The above procedure can be formally justified by the fact that the SABJ model is Markovian, which means that the propagator encodes all the relevant physical information about the process. Markovianity, in conjunction with time-translational invariance, allows us then to interpret one trajectory as an esemble of independent `events', each consisting on the generation of new values for the increments of the process, with each outcome being fully characterized by the same probabilistic model $K$. 

It is important to emphasize that our method can be used to infer parameters from tracking data using generic models and not only the one we consider here. In particular, our method does not depend on the precise shape of the flow profile as far as it is stationary.
It all comes down to computing the corresponding propagator and following the general steps sketched above.
In the next section we discuss the case of the SABJ model and derive the ML estimators as well as their corresponding uncertainties. 

\section{Maximum likelihood method}
\label{sec:ML}

\subsection{Approximate log-likelihood and maximum likelihood estimators}

As discussed above, we need access to the propagator $K$ in order to build the log-likelihood. However, solving the Fokker-Planck equation in presence of a generic flow seems hopeless. This implies that, in practice, the likelihood cannot be exactly constructed. Propitiously, a fairly good approximant can be derived by relying on the relatively large value of experimentally accessible sampling frequencies.
The scheme developed in this section does not rely upon a direct approximate solution of the Fokker-Planck equation for particular flow profiles, since such approach would hinder the generality of our method. As we discuss below, we start from the expression of the continuous-time path probability representation of the SABJ model in presence of a generic flow profile and then introduce time-discretization in the simplest and more natural way possible.
  
To  begin with, let us  first consider the (ideal) limit at which a swimmer  trajectory is sampled continuously, which corresponds to $f\to\infty$. In this case the track probability~\eqref{track-proba} becomes the continuous path probability associated to the process~\eqref{eq-motion-r}-\eqref{eq-motion-p}, which can be written in the Martin-Siggia-Rose-de Dominics-Janssen (MSRDJ) representation (see Refs.~\cite{MSR, DD, Janssen1976} for details) as follows:  
\begin{equation}
\label{path-1}
\M P\big[\Gamma(\bullet)\big]=\M J\big[\Gamma(\bullet)\big]\int\M D\big[\h \Gamma(\bullet)\big]e^{-\h{\M A}[\Gamma(\bullet),\h \Gamma(\bullet)]},
\end{equation}
where $\h \Gamma=(\h{\bo r},\h{\bo p})$ denote the response fields of the MSRDJ formalism, while the action has the form
\begin{equation}
\label{path-2}
\h{\M A}=\int_0^t\bigg\{\ti\big[\dot{\bo r}-\bo v\big]^{\text{T}}\h{\bo r}+\ti\big[\dot{\bo p}-\bo h\big]^{\text{T}}\h{\bo p}+D_R\h{\bo p}^{\text{T}}\tl{\mb M}\h{\bo p}\bigg\}dt'.
\end{equation}

The vector fields $\bo v$ and $\bo h$, as well as the matrix $\tl{\mb M}(\bo p)$, were introduced in Sec.~\ref{sec:prem} and we have omitted their explicit dependence on $\bo r$ and $\bo p$ for compactness. Additionally, the prefactor $\M J$ in Eq.~\eqref{path-1} is a Jacobian whose precise form depends on the underlying discretization scheme of both, the stochastic dynamics~\eqref{eq-motion-r}-\eqref{eq-motion-p} and the path probability. In the Ito convention (which we use here), one has $\M J=1$ and this factor can be omitted.

Imagine now that the sampling frequency is large enough but finite. Then, to first order in $f^{-1}$, one can approximate the time integral involved in the expression of the continuous-time dynamical action~\eqref{path-2} by a discrete sum (with a prescribed discretization scheme, i.e., Ito), which is no more than approximating a Riemann integral by one of its associated Darboux sums when the time step is small enough. 
After integrating over the response variables in~\eqref{path-1}, one gets for the (now discrete) path probability
\begin{equation}
\label{explanation-step-1}
\M P\big[\big\{\Gamma\big\}\big]=\prod_{\alpha=0}^{N-1}\exp\bigg(-\frac{1}{f}\M L(\Gamma(t_{\alpha+1}),\Gamma(t_\alpha))\bigg).
\end{equation}

The precise form of the function $\M L$ is rather involved and we prefer not to discuss it here. In Appendix~\ref{app:likelihood} we develop the full computation. The point we do wish to highlight is that a direct comparison between Eqs.~\eqref{track-proba} and~\eqref{explanation-step-1} explicitly illustrates the nature of our approximation for the propagator $K$ between two events sampled within a very small time interval. In the same vein, taking the logarithm in~\eqref{explanation-step-1} and comparing the result with Eq.~\eqref{log-likelihood} leads to our approximation for the log-likelihood:
\begin{equation}
\label{explanation-step-2}
\M S=-\frac{1}{f}\sum_{\alpha=0}^{N-1}\M L(\Gamma(t_{\alpha+1}),\Gamma(t_\alpha)).
\end{equation}

Following this program, we derive an explicit expression for the log-likelihood of the SABJ model in Appendix~\ref{app:likelihood}. 
The result reads:
\begin{equation}
\label{likelihood-final}
\M S(\tdir,\beta)=N\ln\tdir-\frac{\tdir}{4}\big(A_R-2B_R\beta+C_R\beta^2\big),
\end{equation}
where $\tdir=D_R^{-1}f$ is the inverse of the rotational diffusion coefficient adimensionalized by the sampling frequency. The dimensionless constants $A_R$, $B_R$, and $C_R$ are extensive in the number of sample points $N$ of the trajectory for $N\gg1$ (see Eq.~\eqref{constants}), and depend on the registered position and orientation vectors along the track.

Although the computation leading to~\eqref{likelihood-final} is tedious, the final result exhibits a very simple dependence on both, $D_R$ and $\beta$. Maximizing $\M S$ is now trivial. Provided that $\tdir\neq0$, we obtain the following expressions
\begin{align}
\label{ML-beta}
\beta &=\frac{B_R}{C_R},\\
\label{ML-diff}
\tdir &=\frac{4NC_R}{A_RC_R-B_R^2}.
\end{align}

For consistency, these results must be positive numbers. We discuss below on the accuracy of such estimates.

\subsection{Note on convergence and uncertainties}
\label{sec:accuracy}

From a methodological point of view, it is crucial to be able to quantify the quality of the estimates~\eqref{ML-beta} and~\eqref{ML-diff}. A first important question is whether our approximation for the log-likelihood is compatible with experimental data or not. In other words, we need to provide a quantitative meaning to the statement ``$f$ is sufficiently large''. 
A good way to answer this question is to have a closer look to a consistency condition derived in Appendix~\ref{app:likelihood}, namely (see Eq.~\eqref{Sigma} and the discussion following it) 
\begin{equation}
\label{norm-conservation}
\bo p(t_\alpha)\cdot\Delta\bo p(t_\alpha)+2D_Rf^{-1}|\bo p(t _\alpha)|^2=0,
\end{equation}
where $\Delta\bo p(t_\alpha)$ is the increment of the orientation vector in a time step, $\Delta\bo p(t_\alpha)=\bo p(t_{\alpha+1})-\bo p(t_\alpha)$. This condition is linked to the conservation of the norm of $\bo p$. Indeed, when $f\to\infty$, Eq.~\eqref{norm-conservation} becomes $\bo p(t)\cdot d\bo p(t)=0$. However, an inconsistency may appear if $f$ is not large enough.
To understand this, let us consider the scalar product $\bo p(t_\alpha)\cdot\bo p(t_{\alpha+1})$ for vectors with constant norm, $|\bo p(t_\alpha)|=|\bo p(t_{\alpha+1})|=1$ for all $\alpha$, as is the case for the orientation vector. From one side, we have $\bo p(t_{\alpha})\cdot\bo p(t_{\alpha+1})=\cos(\phi_\alpha)$, where $\phi_\alpha$ is the angle between $\bo p(t_{\alpha})$ and $\bo p(t_{\alpha+1})$. On the other hand, we can write
$\bo p(t_{\alpha})\cdot\bo p(t_{\alpha+1})=\bo p(t_{\alpha})\cdot(\bo p(t_{\alpha})+\Delta\bo p(t_{\alpha}))\equiv1+\bo p(t_{\alpha})\cdot\Delta\bo p(t_{\alpha})$. Using then the condition~\eqref{norm-conservation}, we finally get $\bo p(t_{\alpha})\cdot\bo p(t_{\alpha+1})=1-2D_Rf^{-1}$, which implies $\cos(\phi_\alpha)=1-2D_Rf^{-1}$. It is now clear that our approximation breaks down when $f<D_R$, since in that case, one would have $|\cos(\phi_\alpha)|>1$. In other words, our approximation is not reliable when the rotational diffusion coefficient to be estimated is larger than the sampling frequency.

\begin{figure*}[t!]
	\centering
	\includegraphics[scale=0.45]{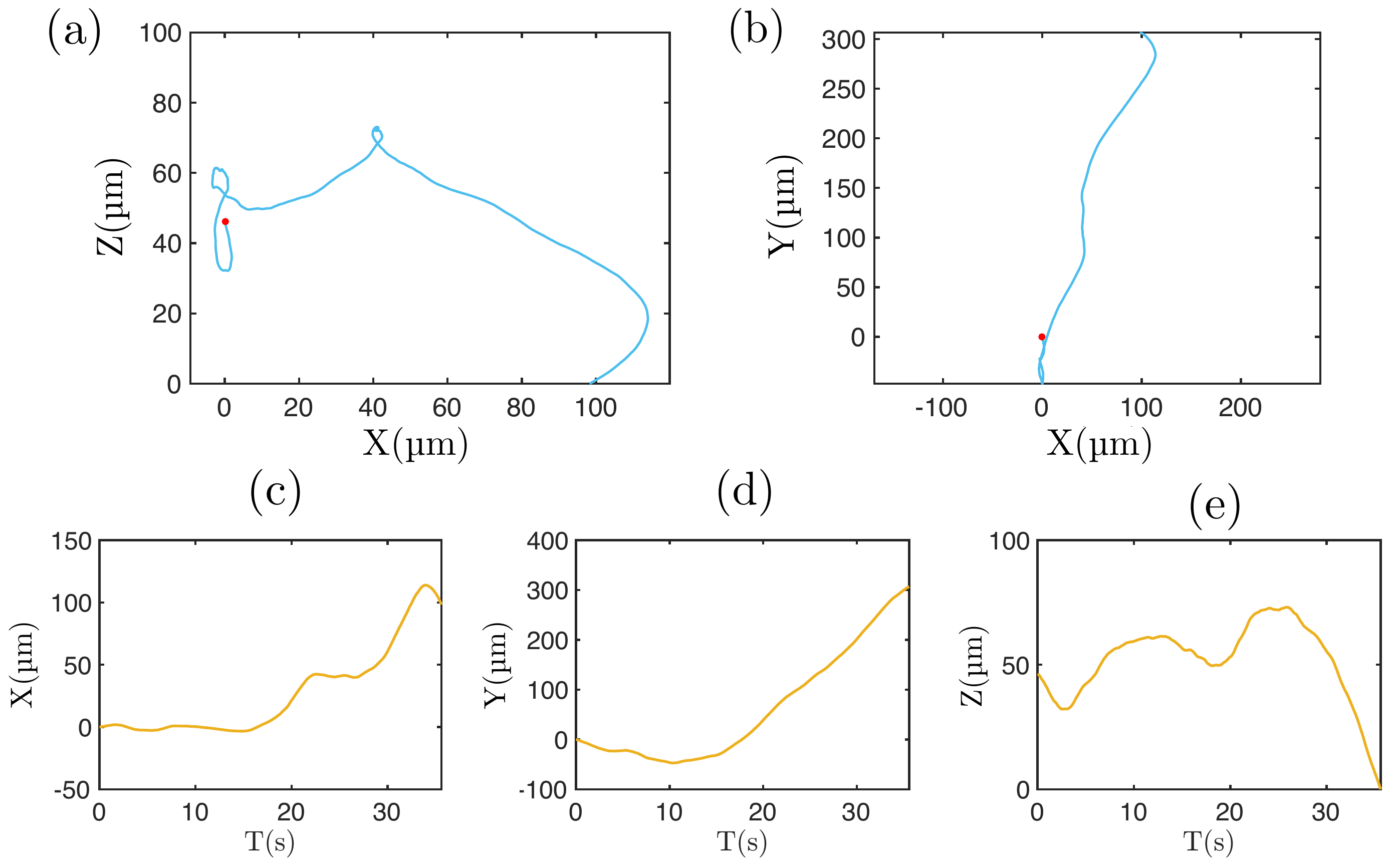}
	\vspace{0 mm}
	\caption{Simulated track of a bacterium swimming at constant speed $v_0=\SI{25}{\micro\meter\per\second}$ in a Poiseuille flow. The height of the channel is $h=\SI{100}{\micro\meter}$. The maximal flow velocity is $U_m = \SI{25}{\micro\meter\per\second}$ along the X direction, which sets $Pe = 100$. The geometric parameter $\beta=0.9$. (a) Projection of the trajectory in the shear plane z-x. (b) Projection of the trajectory in the bottom of the channel i.e., the plane y-x. The red dot indicates the starting point of the trajectory. (c-e) x,y,z coordinates as function of time.}
	\label{Fig_example_trajectory}
\end{figure*}

To assess the efficiency of the method, we estimate the speed at which the computed parameters converge toward their exact values as a function of the length of the sample, $N$. Let us denote by $\bos \theta=(\beta,\tdir)$ the estimated two-dimensional parameter vector, and by $\Delta\bos \theta$ the estimation error. We also write $\bos \theta^{\star}$ to denote the vector of the true values of the parameters. If the assumption that the data is generated by the model~\eqref{eq-motion-r}-\eqref{eq-motion-p} is statistically correct, then one can assert that, up to order $N^{-1/2}$, the estimated parameters are unbiased. More precisely, the error vector $\sqrt{N}\Delta\bos \theta$ converges in distribution to a normal vector of zero mean and variance matrix $\mb F^{-1}(\bos \theta^\star)$, where $\mb F(\bos \theta^\star)$ is the Fisher information matrix~\cite{NEWEY19942111}, which can be approximately determined from the log-likelihood (for large $N$) as $F_{ij}(\bo \theta^\star)=-N^{-1}\partial_{\theta_i,\theta_j}^2S\big|_{\bos{\theta}=\bos{\theta}^\star}$.
Beyond order $N^{-1/2}$, the estimates~\eqref{ML-beta} and~\eqref{ML-diff} are known to exhibit a bias of order $O(1/N)$, but we can neglect such contributions if $N$ is large enough. From this calculation, we estimate the following error bars (we use the notation $e(\theta_i)=\sqrt{\langle\theta_i^2\rangle}$):
\begin{align}
\label{error-beta}
e(\beta) &\sim\frac{2\sqrt{N}}{\sqrt{\tdir^\star[2NC_R-\tdir^\star(B_R-\beta^\star C_R)^2]}},\\
\label{error-diff}
e(\tdir) &\sim\frac{\sqrt{2C_R}\,\tdir^\star}{\sqrt{2NC_R-\tdir^\star(B_R-\beta^\star C_R)^2}}.
\end{align}

Interestingly, there are also cross-correlations between the errors, meaning that $\beta$ and $\tdir$ cannot be determined independently with arbitrarily high accuracy. Explicitly, we have
\begin{equation}
\label{error-cross}
\langle\Delta\beta\Delta\tdir\rangle\sim\frac{2\tdir^\star(B_R-\beta^\star C_R)}{2NC_R-\tdir^\star(B_R-\beta^\star C_R)^2}.
\end{equation}

As a consistency check, note that as $A_R$, $B_R$ and $C_R$ are extensive for $N\gg1$, one has $e(\bos \theta)\sim N^{-1/2}$, as it should be.

\section{Numerical validation of the method}
\label{sec:validation}

In the previous section, we derived ML estimators for the parameters of the SABJ model. 
To test theses expressions and their accuracy, we apply the ML method on simulated bacterial tracks generated using Eqs.~\eqref{eq-motion-r}-\eqref{eq-motion-p}, with an Euler-Maruyama scheme and with a given set of parameters $\beta^*$ and $D_R^*$. Then, we apply the ML method (Eq.~\eqref{ML-beta} and Eq.~\eqref{ML-diff}) on these trajectories to find the estimates $\beta$ and $D_R$. Finally, we compare our estimators with the input parameters of the simulations. 
We study three different cases, namely: (i) a free swimmer (in which case the parameter $\beta$ plays no role), (ii) a simple shear flow (Fig.\ref{Fig_coord_BJ_v3}(c)) and (iii) a Poiseuille flow (Fig.\ref{Fig_coord_BJ_v3}(d)). In Fig.~\ref{Fig_example_trajectory} we illustrate a simulated trajectory of a swimmer in a Poiseulle flow.

\subsection{Free swimmer}

In this case, the orientation of the bacteria follows a diffusion process in the 2-dimensional unit sphere:

\begin{equation}
\dot{\bo p} =  -2\bo p+\sqrt{2}\bo p\wedge\bos \xi_R.
\end{equation}

\begin{figure}[h!]
\centering
	\includegraphics[scale= 0.39]{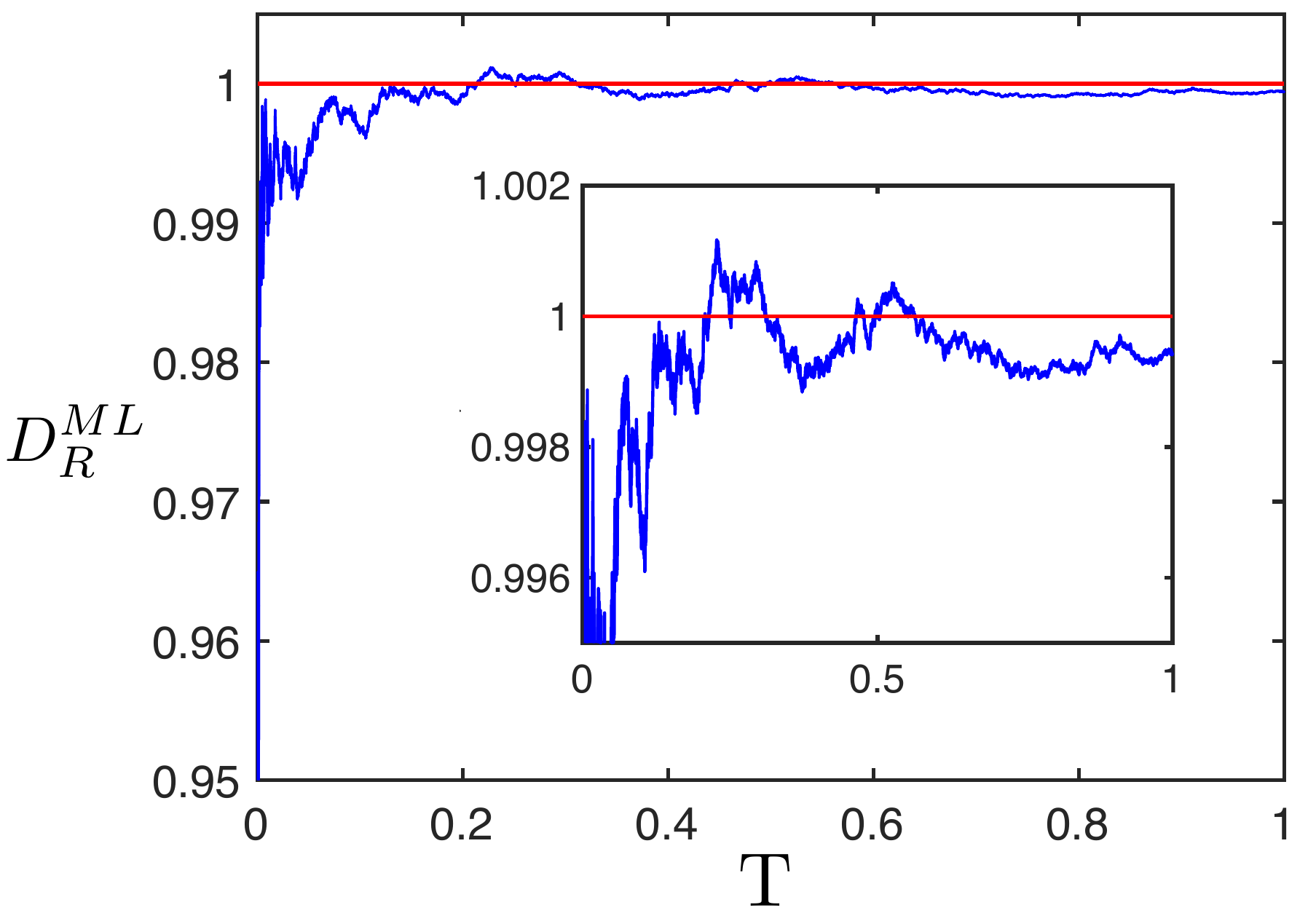}
	\vspace{0 mm}
\caption{Convergence of the estimated rotational diffusion coefficient $D_{R}^{ML}$ in the free swimmer case. Inset: zoom close to the region $D_{R}^{ML}=1$.}
\label{Fig_simple_diffusion}
\end{figure}

We adimensionalized time using the rotational diffusion coefficient $D_R$,
and simulated trajectories with different duration $T$. The (dimensionless) time interval between two sampling events is fixed as $\Delta t=10^{-6}$.
To test the convergence of the method, we compute the diffusion coefficient $D_R^{ML}$, using Eq~\eqref{ML-diff}, for trajectories of different duration.
In Fig. \ref{Fig_simple_diffusion}, we illustrate the convergence of the estimated value $D_{R}^{ML}$ towards the prescribed value $D_R^*=1$ as the length of the trajectory increases.
For instance, we find a value of $D_{R}^{ML}$ very close to 1, i.e $D_{R}^{ML}=0.9994$, at $T=1$.

\subsection{Shear flow}

In this case we simulate Eqs.~\eqref{eq-motion-r} and~\eqref{eq-motion-p} with a flow profile of the form $\bo{v}^F(\bo r) = \dot{\gamma}z\bo{e_x}$,
where $\dot{\gamma}$ is the shear rate.
As before, simulations are performed using dimensionless variables. Time is adimensionalized by the inverse of the shear rate, so the relevant dimensionless, diffusion-related quantity to be estimated in this case is is the P\'eclet number, $Pe=\dot{\gamma}/D_R$.
Note that the P\'eclet number can be directly estimated using Eq.~\eqref{ML-diff}, since it is just a rescaling of $\tdir$ by the known quantities $\dot{\gamma}$ and $f$, $Pe=(\dot{\gamma}/f)\tdir$.

\begin{figure}[hb]
\centering
	\includegraphics[scale= 0.34]{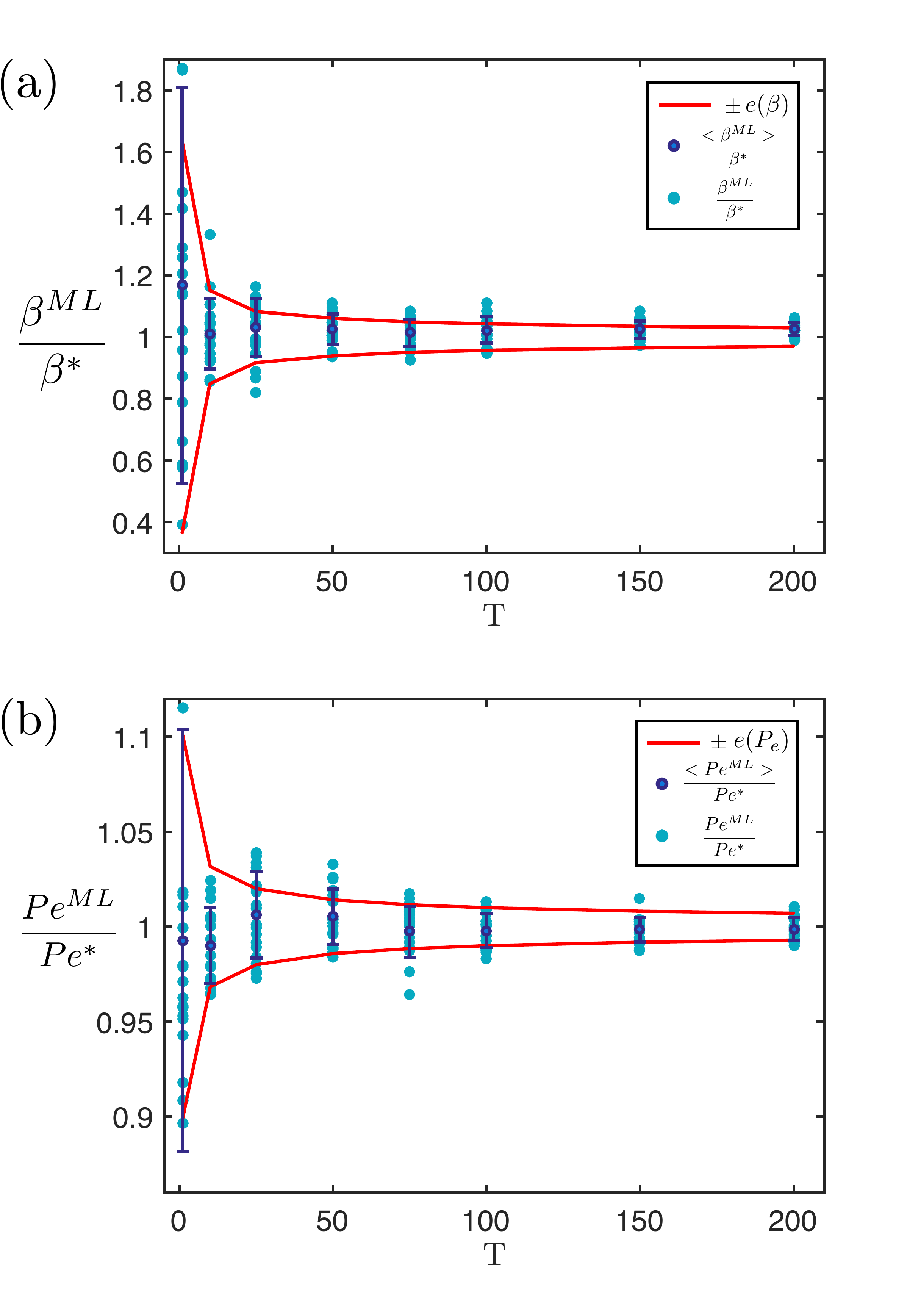}
	\vspace{0 mm}
\caption{Convergence of the estimated (a) $\beta^{ML}$ and (b) P\'eclet $Pe^{ML}$ for simulated trajectories in shear flow. For each duration $T$, 20 trajectories were simulated. Each cyan point corresponds to the estimation of one trajectory, the dark blue points are the average of these estimations over the 20 trajectories. The red curves represent the uncertainties computed using the expressions Eq.~\eqref{error-beta} (resp. Eq.~\eqref{error-diff}).}
\label{Fig_shear_flow}
\end{figure}

We simulate trajectories for a vector $(\beta^* , Pe^*)=(0.9 , 100)$.
To test the convergence of the ML method, we generate a set of trajectories of different duration $T$ with randomly chosen initial positions and orientations.
In Fig.~\ref{Fig_shear_flow}, we present the results of shear flow simulations. 
For a given trajectory length $T$, the estimated value of the parameters ($\beta^{ML}$ in Fig.~\ref{Fig_shear_flow}(a) and $Pe^{ML}$ in Fig.~\ref{Fig_shear_flow}(b)) are scattered around $Pe^*$ and $\beta^*$. This scattering is linked to the randomness encoded in the noise and the initial orientations.
The average value of the estimations over all trajectories of the same length (dark blue points) is very close to the input value of the simulations. On average, even for short trajectories, the ML method gives a good estimation of the parameters. For instance, for a trajectory of duration $T=10$, the averaged estimated parameter differ only by $1 \%$ from the input values. 
As $T$ increases, the scattering of the estimators decreases and our estimators become accurate for individual trajectories. 
Moreover, the standard deviation of the estimated parameters is bounded by the uncertainties computed using the expressions Eq.~\eqref{error-beta} and Eq.~\eqref{error-diff} and decays as $N^{-1/2}$.

\subsection{Poiseuille flow}

\begin{figure}[t!]
\centering
	\includegraphics[scale=0.34]{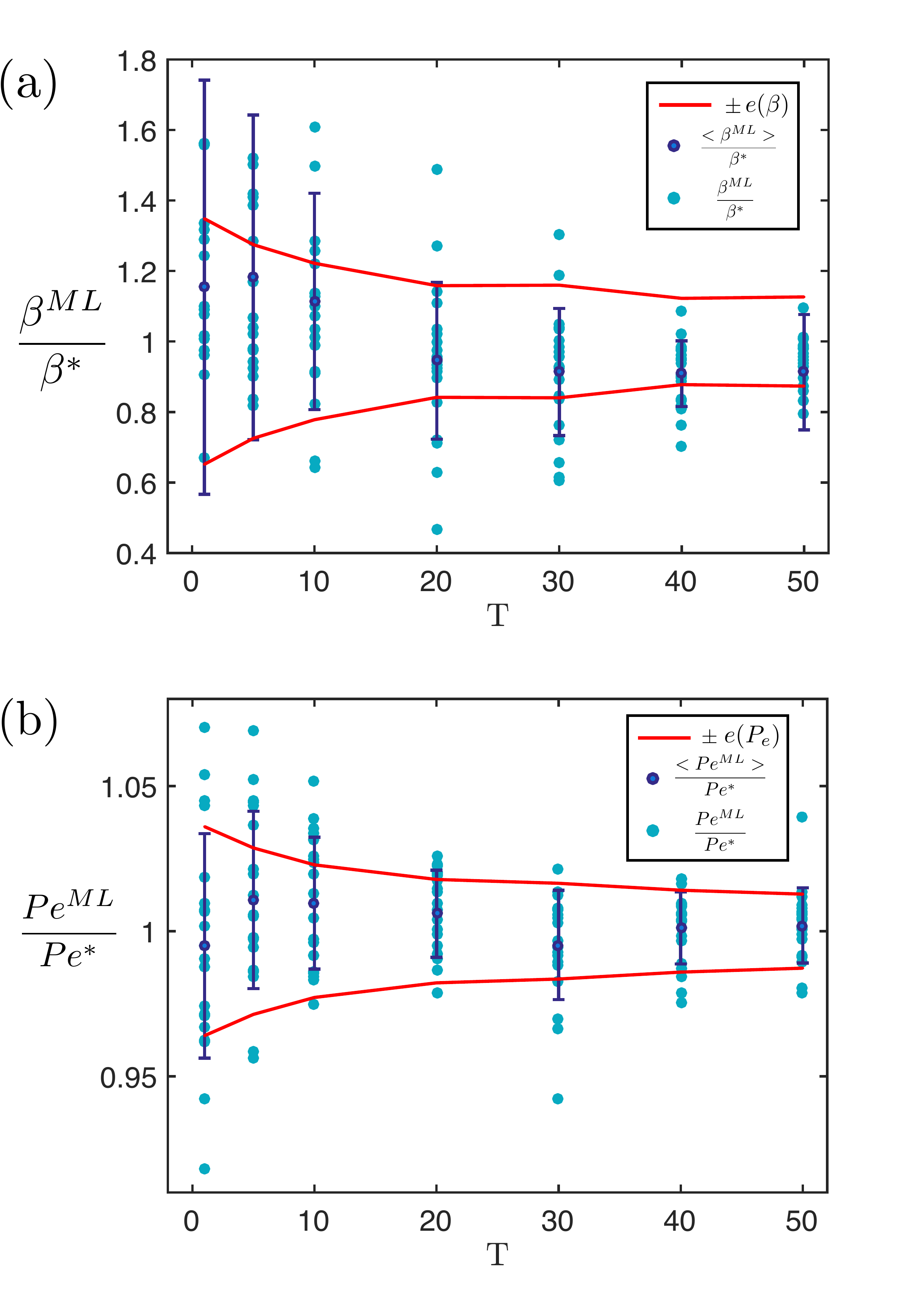}
	\vspace{0 mm}
\caption{Convergence of the estimated (a) $\beta^{ML}$ and (b) P\'eclet $Pe^{ML}$. For each duration $T$, 20 trajectories were simulated. Each cyan point corresponds to the estimation of one trajectory, the dark blue points are the average of these estimations over the 20 trajectories. The red curves represent the uncertainties computed using the expressions Eq.~\eqref{error-beta} (resp. Eq.~\eqref{error-diff}).}
\label{Fig_poiseuille_flow}
\end{figure}

We now test the method on a Poiseuille flow profile. Here we consider a flow confined between two infinite plates separated by a vertical distance $h$, so that the velocity profile is given by $\bo{v}^F(\bo r) = \frac{4U_m}{h}z\Big(1-\frac{z}{h} \Big)\bo{e_x}$.
$U_m$ denotes the maximal flow velocity at the middle of the Poiseille flow ($z=h/2$).
We also work with dimensionless variables in this case, so that lengths and time are adimensionlalized using $h$ and the maximum shear rate $\dot{\gamma}_W=\frac{4U_m}{h}$, respectively. In addition to $\beta$, the relevant dimensionless parameter is again the P\'eclet number, $Pe=\frac{\dot{\gamma}_W}{D_R}$.
We simulate trajectories for a vector: $(\beta^* , Pe^*)=(0.9 , 100)$ and different durations $T$ with the additional constraint that the simulation ends when the particle hits the surfaces (located in z=0 and z=1). For instance, for the parameters used in the Fig.\ref{Fig_poiseuille_flow}, it is difficult to have trajectories longer than T=50.

For a given $T$, the estimators are scattered around the input values. As $T$ increases the scattering decreases. 
The estimator $Pe^{ML}$ remains very good, both on average (less than $2\%$ of error) and for individual trajectories (less than $10\%$ of error), even for trajectories with durations as short as $T=1$.
The estimator $\beta^{ML}$ is not as good as in the shear flow, but we still have an average estimated value within a $10\%$ of the input one.
The standard deviation of the estimated values are still bounded by the expressions Eq.~\eqref{error-beta} and Eq.~\eqref{error-diff} and still decrease as $N^{-1/2}$. Our estimators can be further improved by computing errors up to $O(N^{-1})$ to compensate for the shorter tracks. This will be discussed in more detail elsewhere.

\section{Experimental determination of the parameters}
\label{sec:experimental}

In this section, we give an example of the applicability of the ML method on experimental bacterial trajectories.
A set of Lagrangian 3D swimming trajectories of mutant bacteria with inhibited tumbling (smooth swimmer $\delta$ CheY) was recorded at a frequency $f_{ac}=\SI{80}{\hertz}$ for freely swimming bacteria and $f_{ac}=\SI{100}{\hertz}$ for bacteria in flow.
Further details on the experimental setup are given in the Appendix.

We first consider the free swimmer case. The absence of flow makes the determination of the rotational diffusion coefficient easier and standard methods can be applied. For each trajectory, we computed the correlation function of the orientation vector $\bo p$ as a function of time. It decays exponentially with a characteristic time scale $\tau=(2D_R)^{-1}$, which provides a first estimation of the rotational diffusion coefficient for each bacterium.

To be able to use the ML method, the time interval between two samplings, $\Delta t$ has to fulfill some conditions. As discussed in Sec.~\ref{sec:accuracy}, it has to be small enough so that our approximations are reliable. 
On the other hand, if $\Delta t$ is too small, the change in the bacterium orientation between any instants $t$ and $t+\Delta t$ will be dominated by noise coming from the measurement and not by real physical contributions. 
The optimal value of $f_{ac}$ can be estimated from the theoretical value of the Brownian diffusion coefficient obtained by Perrin \cite{perrin_1934_I,perrin_1936_II}:
\begin{equation}
D_{B} = \frac{3k_BT ln(2l/a)}{\pi \eta l^3}
\label{Dr_brownian}
\end{equation}     
where  $l$ is the length of the ellipsoid and $a$ its width, while $\eta$ denotes de viscosity of the fluid at a temperature $T$.
For an elongated objet of length $l=\SI{10}{\micro\meter}$ and width $a=\SI{1}{\micro\meter}$ (typical bacterium dimensions), it gives $D_B \approx \SI{0.012}{\per\second}$. One should then have $f\gg D_B$; accordingly, we resampled our trajectories at a frequency $1/f=\Delta t=\SI{1/3}{\second}$.

\begin{figure}[t!]
\centering
	\includegraphics[scale=0.36]{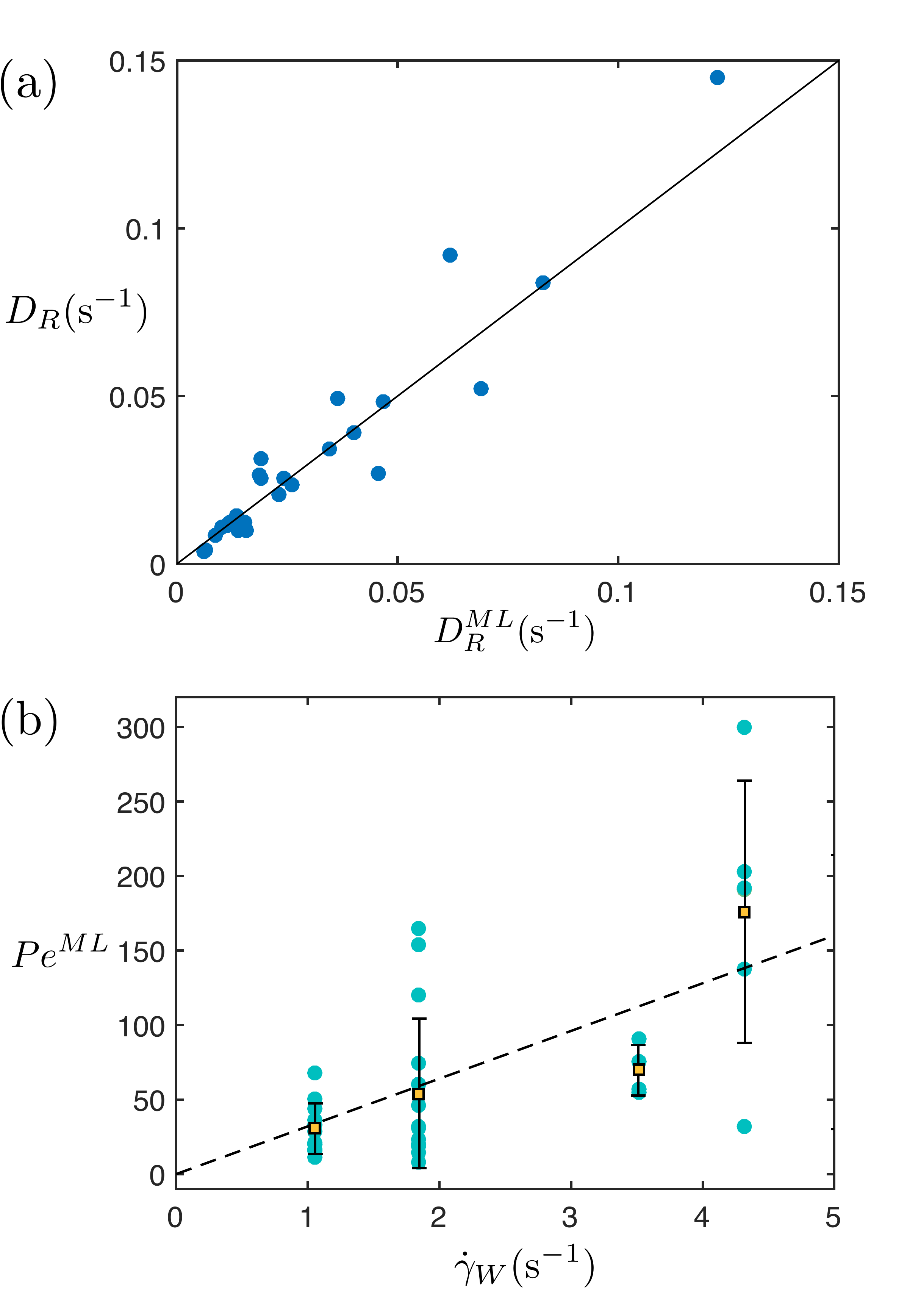}
	\vspace{0 mm}
\caption{(a) Comparison between the rotational diffusion coefficients obtained by the orientation decorrelation method ($D_R$) and with the ML method ($D_R^{ML}$). Each point corresponds to a track. The black line corresponds to the equation $y=x$. Average values are $<D_R>=\SI{0.031}{\per\second}$, with standard deviation $\sigma_R=\SI{0.031}{\per\second}$ and $<D_R^{ML}>=\SI{0.029}{\per\second}$ with standar deviation $\sigma_R^{ML}=\SI{0.027}{\per\second}$. (b) Experimental determination of $Pe^{ML}$. Squares are the average value of all the $Pe^{ML}$ at a given $\dot{\gamma}_W$; the error bars  represent the corresponding standard deviation at a given $\dot{\gamma}_W$. The black dashed line corresponds to a linear fit over the averaged estimated values: $<Pe^{ML}>=\dot{\gamma}_W/D^{ML}$, with $D^{ML} = 0.032 \pm \SI{0.016}{\per\second}$.}
\label{Fig_exp_data}
\end{figure}

By varying the point from where the resampling starts, we obtained from one trajectory $\Gamma(1/f_{ac})$, a set of trajectories $\Gamma_i(1/f)$, $i=1,...,N$, $N=f_{ac}/f$.
We applied the ML method on each resampled trajectory $\Gamma_i$ to obtain $N$ values of the estimated $D_R$ that we averaged to get $D_R^{ML}$; the ML estimation of the rotational diffusion coefficient of the trajectory.
We then applied this method on each of the trajectories.
Fig.~\ref{Fig_exp_data}(a) displays a scatter plot where $D_R$ is in the vertical axis, while the horizontal axis corresponds to $D_R^{ML}$. One can immediately see that both methods give similar results and that the ML estimation works remarkably well. 

In our second experimental test, we injected a suspension of smooth swimmers in a microchannel of rectangular cross section (height $h=\SI{100}{\micro\meter}$ and width $w=\SI{600}{\micro\meter}$). 
Dozens of 3D bacterial trajectories were recorded at different flow rates. 
We focused on bulk trajectories that are at least $\SI{10}{\micro\meter}$ away from the top and the bottom walls and that last at least $10\Delta t$ ($\Delta t$ = \SI{1/3}{\second}). 
The ML method is then used on those trajectories to determine the P\'eclet number.
Fig.~\ref{Fig_exp_data}(b) displays the results of the estimated P\'eclet number $Pe^{ML}$ on experimental bacterial trajectories. 
One can see that $Pe^{ML}$ increases with $\dot{\gamma}_W$, as expected for a constant rotational diffusion coefficient. 
By a linear fit on the average values of $Pe^{ML}$, we get the rotational diffusion coefficient of the bacteria which is the inverse of the slope. It then gives $D^{ML}= 0.032 \pm \SI{0.016}{\per\second}$ which is consistent with $<D_R^{ML}>=\SI{0.029}{\per\second}$ measured in absence of flow, as well as with values reported in presence of flow~\cite{Junot2019}.

\section{Discussion}
\label{sec:discussion}

We have developed an ML method suitable to infer parameters from individual stochastic trajectories.
We were able to estimate the parameters of the SABJ model accurately using both, numerical simulations and experiments.
Preliminary results obtained from the analysis of experimental tracks are encouraging. 
One of the outcomes of our numerical study is that uncertainties fit well an $O(N^{1/2})$ decay even for relatively short trajectories. As an important remark, note such estimates
can in principle be further improved in a systematic way by considering errors up to $O(N^{-1})$. 

Our estimators for the rotational diffusion coefficient are much more scattered in experiments than in simulations.
In particular, standard deviations in Figs.~\ref{Fig_exp_data}(a) and (b) are large. However, it is important to keep in mind that each point in Fig.~\ref{Fig_exp_data} represents a different bacterium, which implies that one should be careful not to conflate biological variability with a lack of accuracy of the method. 
In support of the hypothesis of individual variability, note that the shape of a bacterium affects the value of $D_R$, which depends sensitively on the effective value of the ellipsoid long axis $l$. Such $l^{-3}$ dependence may naturally lead to a rather broad distribution of coefficient values within a population, which is indeed what was found.
For instance, for an effective ellipsoid of length $l=5$ to $\SI{10}{\micro\meter}$ and width $a=\SI{1}{\micro\meter}$, the estimation of $D_R$ using Eq.~\eqref{Dr_brownian}, yields $D_R = 1.25$ to $5\times10^{-3}\SI{}{\per\second}$. We thus conclude that there is an intrinsic variability of $D_R$ within a bacterial population, which is precisely why one needs a method like the one developed in this work to analyze individual trajectories.

The tool we have devised here is general in a natural way and can be applied to any Markovian and time-translationally invariant stochastic process. In the context of swimming bacteria, one can study variations of parameters due to external conditions and physical effects that are not considered in the SABJ model. For instance, shear may shorten the flagella bundle leading to higher effective $D_R$ (i.e lower $Pe$). Modifying the SABJ model as to account for rheotactic effects~\cite{son2013bacteria,sporing2018hook}, and addapting our ML method appropriately, one could study rheotactic drift~\cite{Marcos2012} in detail, analyze possible changes in bundle formation due to shear, and probe the flexibility of the hook at the base of the flagellum and its interplay with external flow. These and other physical questions will be addressed in the immediate future.


\acknowledgments
This work was supported by the ANR grant ``BacFlow'' ANR-15-CE30-0013 and by Institut Pierre-Gilles de Gennes 
(\'Equipement d'Excellence, Investissements d'Avenir
program ANR-10-EQPX-34). RGG acknowledges support fron ANR grant ANR-17-CE08-0047-02.
E.C. is supported by the Institut Universitaire de France. We thank Carine Douarche for providing us the E.coli CR20 smooth mutant. We are also very thankful to the Edinburgh group of Pr Wilson Poon and in particular to Angela Dawson and Vincent Martinez, for providing the 2-color AD63 E.coli strain we display on Fig.1.

 
\newpage 
 
\appendix

\section{Approximate log-likelihood for the SABJ model}
\label{app:likelihood}
 
In this secction we go through the full calculation leading to the result~\eqref{likelihood-final} for the log-likelihood of the SABJ model. Our starting point is the MSRDJ path-integral representation of the process~\eqref{eq-motion-r}-\eqref{eq-motion-p}, which is given by Eqs.~\eqref{path-1} and~\eqref{path-2} with $\M J=1$.
As mentioned in the text, one approximates the time integrals by discrete sums in the action when the sampling frequency is large enough. We write 
\begin{widetext}
\begin{equation}
\label{path-3}
\h{\M A}[\{\Gamma,\h \Gamma\}]=\frac{1}{f}\sum_\alpha\bigg\{\ti\big[f\Delta{\bo r}_\alpha-\bo v_\alpha\big]^{\text{T}}\h{\bo r}_\alpha+\ti\big[f\Delta{\bo p}_\alpha-\bo h_\alpha\big]^{\text{T}}\h{\bo p}_\alpha+D_R\h{\bo p}_\alpha^{\text{T}}\tl{\mb M}(\bo p_\alpha)\h{\bo p}_\alpha \bigg\},
\end{equation}
where $\Gamma_\alpha$ ($\h \Gamma_\alpha$) is a shorthand notation for $\Gamma(t_\alpha)$ ($\h \Gamma(t_\alpha)$), while $\Delta\Gamma_\alpha\equiv\Gamma_{\alpha+1}-\Gamma_\alpha$.
Additionally, we have introduced the notations $\bo v_\alpha\equiv\bo v(\Gamma_\alpha)$ and $\bo h_\alpha\equiv\bo h(\Gamma_\alpha)$. With all this, we write
\begin{equation}
\label{path-4}
\M P[\{\bo r,\bo p\}]=\int\prod_\alpha d^3\h{\bo r}_\alpha d^3\h{\bo p}_\alpha\exp\bigg(-\frac{1}{f}\sum_\alpha\bigg\{\ti\big[f\Delta{\bo r}_\alpha-\bo v_\alpha\big]^{\text{T}}\h{\bo r}_\alpha+\ti\big[f\Delta{\bo p}_\alpha-\bo h_\alpha\big]^{\text{T}}\h{\bo p}_\alpha+D_R\h{\bo p}_\alpha^{\text{T}}\tl{\mb M}(\bo p_\alpha)\h{\bo p}_\alpha \bigg\}\bigg).
\end{equation}

The integrals over $\{\h{\bo r}_\alpha\}$ are immediate and can be readily performed to yield
\begin{equation}
\label{path-5}
\M P[\{\bo r,\bo p\}]=\prod_\alpha\delta\big(\Delta{\bo r}_\alpha-f^{-1}\bo v_\alpha\big)\int\prod_\alpha d^3\h{\bo p}_\alpha\exp\bigg(-\frac{1}{f}\sum_\alpha\bigg\{\ti\big[f\Delta{\bo p}_\alpha-\bo h_\alpha\big]^{\text{T}}\h{\bo p}_\alpha+D_R\h{\bo p}_\alpha^{\text{T}}\tl{\mb M}(\bo p_\alpha)\h{\bo p}_\alpha \bigg\}\bigg).
\end{equation} 

The remaining integrals over $\{\h{\bo p}_\alpha\}$ are less trivial because, as can be verified, the matrix $\tl{\mb M}$ is singular, i.e., $\det(\tl{\mb M})=0$. Such singularity is linked to the fact that the continuous dynamics~Eq.~\eqref{eq-motion-p} imposes a hard constraint on $\bo p$, i.e., $|\bo p|^2=1$. This means that at any time instant only two numbers are needed to fully describe the state of the three-dimensional vector $\bo p$. 
For instance, by knowing the projection of $\bo p$ on the $z$ axis, $p_z$, and the polar angle $\theta$ of the projection of $\bo p$ in the $x-y$ plane, one can write $\bo p=(\sqrt{1-p_z^2}\cos\theta,\sqrt{1-p_z^2}\sin\theta,p_z)$.

From this discussion, one can infer that there is a relevant two-dimensional sub-space allowing to fully describe $\bo p$. One can indeed isolate the singularity, so that a Gaussian integration can still be made in a reduced, two-dimensional submanifold. To show this, we introduce the change of variables $\h{\bo{p}}_\alpha=\mb R_\alpha\bos \omega_\alpha$, where the matrix $\mb R_\alpha$ is given as
\begin{equation}
\label{R-matrix}
\mb R_\alpha=\begin{bmatrix}
p_{x,\alpha} & -p_{z,\alpha} & -\frac{p_{y,\alpha}p_{x,\alpha}^2}{p_{x,\alpha}^2+p_{z,\alpha}^2}\\
\\
p_{y,\alpha} & 0 & p_{x,\alpha}\\
\\
p_{z,\alpha} & p_{x,\alpha} & -\frac{p_{x,\alpha}p_{y,\alpha}p_{z,\alpha}}{p_{x,\alpha}^2+p_{z,\alpha}^2}
\end{bmatrix}.
\end{equation}

One can check that the determinant of $\mb R_\alpha$ is given by $\det(\mb R_\alpha)=-p_{x,\alpha}|\bo p_\alpha|^2$. Introducing the new vector 
\begin{equation}
\label{rho-vec}
\bos \rho_\alpha=\mb R_\alpha^{\text{T}}[f\Delta{\bo p}_\alpha-\bo h_\alpha],
\end{equation}
we have:
\begin{equation}
\label{path-6}
\M P[\{\bo r,\bo p\}]=\prod_\alpha\delta\big(\Delta{\bo r}_\alpha-f^{-1}\bo v_\alpha\big)I[\{\bo r,\bo p\}],
\end{equation}
where
\begin{equation}
\label{path-7}
I[\{\bo r,\bo p\}]=\int\prod_\alpha d^3\bos{\omega}_\alpha\exp\bigg(-\frac{1}{f}\sum_\alpha\bigg\{\ti\bos \rho_\alpha^{\text{T}}\bos \omega_\alpha+D_R\bos \omega_\alpha^{\text{T}}\mb{\Gamma}_\alpha\bos \omega_\alpha-f\ln\big(|p_{x,\alpha}||\bo p_\alpha|^2\big) \bigg\}\bigg),
\end{equation}
and the matrix $\mb{\Gamma}_\alpha$ has the following form:
\begin{equation}
\label{gamma-matrix}
\mb \Gamma_\alpha=\begin{bmatrix}
0 & 0 & 0\\
\\
0 & (p_{x,\alpha}^2+p_{z,\alpha}^2)|\bo p_\alpha|^2 & 0\\
\\
0 & 0 & \frac{p_{x,\alpha}^2|\bo p_\alpha|^4}{p_{x,\alpha}^2+p_{z,\alpha}^2}
\end{bmatrix}.
\end{equation}

The integrals over $\{\bos \omega_\alpha\}$ in~\eqref{path-7} can now be immediately performed to yield
\begin{equation}
\label{path-8}
I[\{\bo r,\bo p\}]=\bigg(\frac{\pi f}{D_R}\bigg)^N\prod_{\alpha}\bigg[\frac{1}{|\bo p_\alpha|}\delta(f^{-1}\rho_{x,\alpha})\bigg]\exp\bigg(-\frac{1}{4D_Rf}\sum_\alpha\bigg\{\frac{\rho_{y,\alpha}^2}{(p_{x,\alpha}^2+p_{z,\alpha}^2)|\bo p_\alpha|^2}+\frac{\rho_{z,\alpha}^2(p_{x,\alpha}^2+p_{z,\alpha}^2)}{p_{x,\alpha}^2|\bo p_\alpha|^4}\bigg\}\bigg).
\end{equation}

This result can be written in a more illustrative way if one calculates $\rho_{x,\alpha}$ explicitly using~\eqref{rho-vec}. One has $f^{-1}\rho_{x,\alpha}=f^{-1}\bo p_\alpha\cdot[f\Delta\bo p_\alpha-\bo h_\alpha]\equiv \bo p_\alpha\cdot\Delta\bo p_\alpha+2D_Rf^{-1}|\bo p _\alpha|^2$. With all this we finally obtain that the probability of a given sequence $\{\bo r,\bo p\}$, in this approximation, is
\begin{equation}
\label{semifinal-path-proba}
\M P[\{\bo r,\bo p\}]=\bigg(\frac{\pi f}{D_R}\bigg)^N\Sigma[\{\bo r,\bo p\}]\exp\bigg(-\frac{1}{4fD_R}\sum_\alpha\bigg\{\frac{\rho_{y,\alpha}^2}{(p_{x,\alpha}^2+p_{z,\alpha}^2)|\bo p_\alpha|^2}+\frac{\rho_{z,\alpha}^2(p_{x,\alpha}^2+p_{z,\alpha}^2)}{p_{x,\alpha}^2|\bo p_\alpha|^4}+4fD_R\ln|\bo p_\alpha|\bigg\}\bigg),
\end{equation}
where $\Sigma[\{\bo r,\bo p\}]$ is a singular measure inforcing a set of constraints on the trajectories generated by Eqs.~\eqref{eq-motion-r} and~\eqref{eq-motion-p}:
\begin{equation}
\label{Sigma}
\Sigma[\{\bo r,\bo p\}]=\prod_{\alpha}\delta\big(\Delta{\bo r}_\alpha-f^{-1}\bo v_\alpha\big)\prod_{\alpha}\delta\big(\bo p_\alpha\cdot\Delta\bo p_\alpha+2D_Rf^{-1}|\bo p _\alpha|^2\big).
\end{equation}

Let us analyze these constraints in more detail. The condition $\bo p_\alpha\cdot\Delta\bo p_\alpha+2D_Rf^{-1}|\bo p _\alpha|^2=0$ is related the conservation of the norm of $|\bo p|$. Note that, in particular, when $f\to\infty$, it takes the form $\bo p(t)\cdot d\bo p(t)=0$, as one would expect. On the other hand, the condition $\Delta{\bo r}_\alpha-f^{-1}\bo v_\alpha=0$ provides, in practice, a practical way to determine $p_\alpha$ at each time step. Indeed, the tracking device measures the trajectory of the particle, while $p$ has to be determined indirectly. Writing this condition more explicity as $\Delta{\bo r}_\alpha-f^{-1}(v_0\bo p_\alpha+\bo v^F_\alpha)=0$, one has $\bo p_\alpha=v_0^{-1}(f\Delta{\bo r}_\alpha-\bo v^F_\alpha)$, which is the expression that is used in practice to determine $\bo p_\alpha$ at each time step. In summary, if we restrict our analysis to the relevant sub-space of the trajectories which are compatible with the constraints $\Sigma$ (inforcing, in particular, that $|\bo p_\alpha|=1$ for all $\alpha$), we have the following expression for the discretized path probability:
\begin{equation}
\label{final-path-proba}
\M P_\Sigma[\{\bo r,\bo p\}]=\bigg(\frac{\pi f}{D_R}\bigg)^N\exp\bigg(-\frac{1}{4fD_R}\sum_\alpha\bigg\{\frac{\rho_{y,\alpha}^2}{p_{x,\alpha}^2+p_{z,\alpha}^2}+\frac{\rho_{z,\alpha}^2(p_{x,\alpha}^2+p_{z,\alpha}^2)}{p_{x,\alpha}^2}\bigg\}\bigg).
\end{equation}

To proceed further, we introduce the notations $\bo a_\alpha=f\Delta\bo p_\alpha-(\mb 1-\bo p_\alpha\otimes\bo p_\alpha)\mb \Omega(\bo r_\alpha)\bo p_\alpha$, and $\bo b_\alpha=(\mb 1-\bo p_\alpha\otimes\bo p_\alpha)\mb E(\bo r_\alpha)\bo p_\alpha$, so that we have $f\Delta\bo p_\alpha-\bo h_\alpha\equiv \bo a_\alpha-\beta\bo b_\alpha+2D_R\bo p_\alpha$. We also define
\begin{align}
\label{constants}
S_\alpha &=p_{x,\alpha}^2+p_{z,\alpha}^2\nonumber\\
A_\alpha &=a_{z,\alpha}p_{x,\alpha}-a_{x,\alpha}p_{z,\alpha},\nonumber\\
B_\alpha &=b_{z,\alpha}p_{x,\alpha}-b_{x,\alpha}p_{z,\alpha},\nonumber\\
C_\alpha &=a_{y,\alpha}S_\alpha-(a_{x,\alpha}p_{x,\alpha}+a_{z,\alpha}p_{z,\alpha})p_{y,\alpha},\nonumber\\
E_\alpha &=b_{y,\alpha}S_\alpha-(b_{x,\alpha}p_{x,\alpha}+b_{z,\alpha}p_{z,\alpha})p_{y,\alpha},\nonumber\\
A_R &=f^{-2}\sum_{\alpha}S_\alpha^{-1}(A_\alpha^2+C_\alpha^2),\nonumber\\
B_R &=f^{-2}\sum_{\alpha}S_\alpha^{-1}(A_\alpha B_\alpha+C_\alpha E_\alpha),\nonumber\\
C_R &=f^{-2}\sum_{\alpha}S_\alpha^{-1}(B_\alpha^2+E_\alpha^2).
\end{align}

The log-likelihood can be derived by taking the logarithm of Eq.~\eqref{final-path-proba} after having into account that $|\bo p_\alpha|=1$ for all $\alpha$ in any experimental or numerical track, and that Eq.~\eqref{eq-motion-r} is used to fix the evolution of the vector $\bo p$. To take the logarithm, we define $\tl{\M P}=\pi^{-N}\M P_\Sigma$, with $\M P_\Sigma$ given by~\eqref{final-path-proba}. It is also convenient to introduce the dimensionless inverse diffusion coeffient $\tdir=D_R^{-1}f$. In terms of the quantities introduced in~\eqref{constants}, we write $S=\ln\tl{\M P}$, which gives Eq.~\eqref{likelihood-final}. 	
\end{widetext}

\section{Experimental set-up and protocol}
\label{app:exp}

Bacteria used in this work are smooth swimmer mutants of an \ecoli (strain CR20, $\Delta$-CheY) that almost never tumble and were transformed with a plasmid coding for a yellow fluorescent protein (YFP). Bacteria are grown overnight at 30$^\circ$C until the early stationary phase.
The growth medium is then removed by centrifuging the culture and removing the supernatant. The bacteria are resuspended in a Motility Buffer (MB) below the very low concentration of $3\times10^7$ bacteria per mL, in order to visualize one bacterium at a time and to minimize the interactions between bacteria. The suspension is supplemented with L-serine at 0.08g/mL and polyvinyl pyrrolidone (PVP) at 0.005$\%$; L-serine maintains good motility for a few hours and PVP is used to prevent bacteria from sticking to surfaces. The solution is also mixed with Percol (1:1) to avoid bacterial sedimentation. The experiments are performed at a temperature of $25^{o}$C. 

The channel is visualized using a home-made Lagrangian tracking microscope \cite{Darnige2016} here used to track fluorescent swimming bacteria. By a visualization based feedback acting on a mechanical (horizontal) and piezoelectric (vertical) stage, the targeted object is kept close to the center of the visualization field and in focus on an inverted microscope (Zeiss-Observer, Z1 with an objectif C-Apochromat 63x/1.2 W). Images of the tracked objects are acquired at 80 or \SI{100}{\hertz} with a Hamamatsu Orca-flash 4.0 camera. Simultaneously, the three-dimensional positions of the object are recorded.

The measurements take place in a microfluidic channel of rectangular cross-section (height $h = 100 ~\mu m$, width $w = 600 ~\mu m$), made in Polydimethylsiloxane (PDMS) using standard soft-lithography techniques. Flow is imposed through the channel via a Nemesys syringe pump (dosing unit Low Pressure Syringe Pump neMESYS 290N and base Module BASE 120N). The flow rate varies from 1 to \SI{4.3}{\nano\liter\per\second} corresponding respectively to $U_{m}$ between $(28 \pm1.9)~\mu \text{m/s}$ and $(120 \pm4.0)~\mu \text{m/s}$ and maximal shear rates $\dot{\gamma}_M=4 U_{m}/h$ between $(1.12 \pm0.076)~\text{s}^{-1}$ and $(4.82\pm0.16)~\text{s}^{-1}$. 
We set our region of interest in the center of the channel with respect to its width and consider only trajectories which are at least 200$\mu m$ away from the lateral walls.

\bibliographystyle{apsrev4-1.bst}

\end{document}